# Long-wave infrared Fourier transform spectroscopy with enhanced and scalable sensitivity

SERGEY VASILYEV,[1,†,*] RODERIK KREBBERS,[2,†] DMITRII KONNOV,[4] MATHIEU WALSH,[3] IGOR MOSKALEV,[1] MIKE MIROV,[1] ANDREY MURAVIEV,[4] JÉRÔME GENEST,[3] KONSTANTIN VODOPYANOV[4] AND SIMONA M. CRISTESCU[2]

[1]*IPG Photonics Corporation, 377 Simarano Dr, Marlborough, MA 01752, USA*
[2]*Institute for Molecules and Materials, Radboud University, Heyendaalseweg 135, 6525 AJ Nijmegen, Netherlands*
[3]*Centre d'Optique, Photonique et Laser, Université Laval, Québec, Quebec G1K 7P4, Canada*
[4]*CREOL, the College of Optics and Photonics, University of Central Florida, Orlando, Florida 32816, USA*
*† These authors contributed equally to this work.*
*\*svasilyev@ipgphotonics.com*

**Abstract:** We report a broadband long-wave infrared Fourier transform spectrometer with sensitivity exceeding that of previously reported direct-detection implementations. The system combines dual-comb spectroscopy with electro-optic sampling, multi-channel parallel near-infrared detection using InGaAs photodiodes, and real-time GPU-based computational corrections of multiple spectroscopy signals. Detection limits of 0.3 ppb for $NH_3$ and 2 ppb for $C_2H_4$ are achieved in 500 s, corresponding to 20× and 40× sensitivity improvements over earlier LWIR demonstrations, while maintaining high 0.0027 cm$^{-1}$ spectral resolution and broad spectral coverage. The architecture supports scalable sensitivity through increased detector count and enables rapid multispecies analysis of complex gas mixtures.

## 1. Introduction

Fourier transform spectroscopy (FTS) is a well-established technique that is widely used in science, medicine, and industry. A large fraction of FTS applications target the long-wave IR spectral range (LWIR, 5 – 20 μm or 500 – 2000 cm$^{-1}$) – the so-called molecular fingerprint region, which contains strong, distinctive absorption bands of many molecules. Dual comb spectroscopy (DCS) is a subtype of FTS that retains all the strengths of conventional FTS while adding all the features of laser spectroscopy [1]. The key advantages of the DCS technique are high, kHz-level, spectroscopic resolution, direct linking of the optical frequency scale to an atomic clock, and the ability to acquire individual spectra at high repetition rates. As a result, DCS has become an essential tool for advanced applications such as generating high accuracy line lists for molecular spectroscopic databases [2, 3, 4] and time-resolved spectroscopic measurements of fast processes [5, 6, 7].

Despite these advantages, the sensitivity – one of the most important parameters for real-world applications – of FTS and DCS remains fundamentally equivalent. The two techniques share the same figure of merit (FOM), derived in [8] for DCS and known from [9] for FTS. Consequently, FTS- and DCS-based instruments will yield the same sensitivity if they measure the spectrum of a light source with the same power spectral density (PSD) distribution, at the same spectral resolution, during the same total time, and using the same detector. Figure 1 illustrates different sensitivity regimes of FTS and DCS, where performance is limited by the source's power (circles), source's intensity noise (squares), or detector noise (diamonds). To further enhance the sensitivity, the authors of [8] suggested parallel spectral acquisition with multiple detectors (double diamond). However, as also follows from Fig. 1, any sensitivity enhancement demands a light source with simultaneously higher power and lower intensity noise – a combination of requirements that is difficult to implement in practice, especially in the LWIR. The actual sensitivity is further limited by the phase noise of the frequency combs (in case of DCS), the quality of the interferometer (in case of FTS), and many other practical considerations. For instance, in the LWIR and mid-wave IR (MWIR) regimes, the use of multiple detectors for parallel spectral acquisition must be balanced by the corresponding increase in system complexity. Hence, the number of the photodetectors $N_D$ in practical implementations is limited to $N_D = 2$ [6].

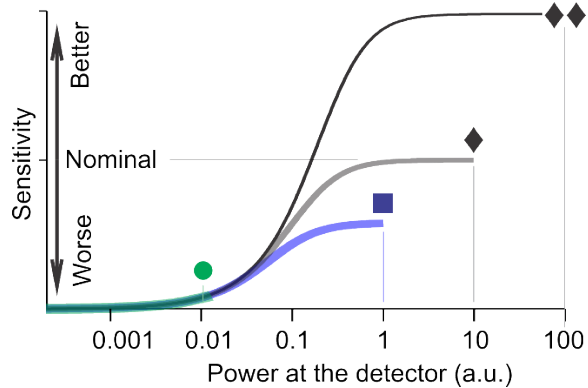

Fig. 1 FTS/DCS sensitivity vs power of the light source in log scale (qualitative estimates based on the formulate derived in [8]): ●, source power limited regime; ■, source noise limited regime, ♦, detector limited regime; ♦♦, sensitivity enhancement via parallel spectral acquisition. Regimes ♦ and ♦♦ require a source with ultra-low intensity noise.

The highest sensitivity of a Fourier spectrometer reported to date was achieved in the MWIR spectral range using cavity-enhanced direct-frequency comb spectroscopy (CE-DFCS) technique. For instance, detection limits of 0.13 ppb, 0.64 ppb, 19 ppb, and 103 ppb were obtained in 200 seconds for formaldehyde ($H_2CO$), methane ($CH_4$), ethylene ($C_2H_4$), and ammonia ($NH_3$), respectively, at the MWIR wavelength 3.5 μm [10]. There also has been a recent report on cavity-enhanced FTS in the LWIR [11]. However, the authors did not assess the sensitivity of their instrument. It is also important to note that the spectral window of a CE-DFCS-based Fourier spectrometer is limited by an interplay between chromatic dispersion and finesse of the enhancement cavity. Therefore, the most sensitive CE-DFCS instruments have relatively narrow spectral windows of about 100 cm$^{-1}$ [12].

In our prior work [13], we reported a highly sensitive and broadband FTS in the MWIR and LWIR using a frequency comb and a simple multi-pass cell. In the MWIR, we achieved a detection sensitivity of 0.2 ppb for methane in 1000 seconds in a spectral window 2860 – 3330 cm$^{-1}$, thus outperforming a sophisticated CE-DFCS instrument (albeit at a lower spectral resolution of 0.1 cm$^{-1}$). In the LWIR, we demonstrated a detection limit of 6.2 ppb ammonia in 500 seconds in a spectral window 900 –1250 cm$^{-1}$. We believe this is the highest FTS sensitivity reported to date in the LWIR.

With this result, we approached the nominal level of FTS performance that is imposed by the dynamic range of available LWIR photodetectors. We attributed this to the use of a high-power, low-noise MWIR/LWIR source based on the ultrafast Cr:ZnS laser technology. In another set of experiments, we used Cr:ZnS-based optical frequency combs to implement LWIR DCS that is simultaneously broadband, high resolution, and fast [14].

Here we push the performance envelope of Fourier transform infrared spectroscopy to a new regime of enhanced sensitivity, without compromising the instrument's bandwidth and resolution, as summarized in Tables 1 and 2. Specifically, in comparison to the previous record [13], we enhanced sensitivity by an order of magnitude at 40× higher spectral resolution (0.0027 cm$^{-1}$) and in a 2× broader spectral window (800 – 1400 cm$^{-1}$). To achieve this, we combined an improved version of the same laser source with DCS and used the electro-optic sampling technique (EOS-DCS, [15]). Further, we exploited the high power, short pulses, and ultra-low noise of ultrafast Cr:ZnS lasers to implement parallel acquisition of LWIR spectra with marginal increase in system complexity.

**Table 1. Sensitivity of selected spectrometers in the MWIR and LWIR**

| Molecule | Detection limit in 200 seconds, ppb | | |
|---|---|---|---|
| | Ref. [10] (#) | Ref. [13] (#) | This work (#) |
| $CH_4$ | 0.64 (A) | 0.4 (B) | NA |
| $NH_3$ | 103 (A) | 7 (C) | 0.8 (D) |
| $C_2H_4$ | 19 (A) | 100 (C) | 4 (D) |

**Table 2. Parameters of selected spectrometers**

| # | Comb source | Spectral window, $cm^{-1}$ | Spectral resolution, $cm^{-1}$ | Number of spectral elements | Interaction length, m |
|---|---|---|---|---|---|
| A | Yb:fiber laser | 2810 – 2945 | 0.00907 | 14,900 | 1300[a] |
| B | Cr:ZnS laser | 2857 – 3333 | 0.1 | 4,800 | 31.2[b] |
| C | Cr:ZnS laser | 900 – 1250 | 0.1 | 3,500 | 31.2[b] |
| D | Cr:ZnS laser | 800 – 1400 | 0.00267 | 224,700 | 31.2[b] |

[a] Effective interaction length in the enhancement cavity with finesse $F = 7500$ and FSR = 272 MHz; [b] multi-pass cell.

## 2. Ultra-low noise Cr:ZnS comb source for high-performance Fourier transform spectroscopy

Cr:ZnS and its sister material Cr:ZnSe belong to a large family of transition-metal-doped II-VI semiconductors [16]. Due to their broad tuning range of 1.8 – 3.3 μm and room temperature (RT) operation, these lasers are often referred to as the "Ti:sapphires of the middle-infrared". In fact, compared to Ti:sapphire, the Cr:ZnS has a higher peak emission cross-section ($14 \cdot 10^{-19}$ $cm^2$ vs. $4 \cdot 10^{-19}$ $cm^2$) and longer RT lifetime (4.3 μs vs. 3 μs) that provides a lot of flexibility in the ultrafast laser design [17].

Another feature of ultrafast polycrystalline Cr:ZnS lasers and amplifiers is pronounced three-wave mixing directly inside the gain elements that occurs due to random quasi-phase-matching (RQPM) in polycrystals [18]. The distinctive features of RQPM are a linear dependence of the conversion yield on the medium length, and an ultra-wide bandwidth, which is well suited for frequency conversion of fs pulses with a high peak power and a broad spectrum.

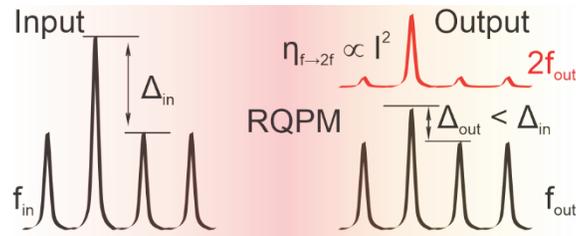

Fig. 2 Intensity noise suppression during the propagation of a femtosecond pulse train at the fundamental wavelength (f) through an RQPM medium. A part of the input pulse train is converted to the second harmonic (2f) due to three-wave mixing in the RQPM medium (the f → 2f conversion efficiency is quadratic on the intensity of pulses) $\Delta_{in}$ and $\Delta_{out}$ are relative amplitudes of an intensity spike in the pulse train before and after the medium.

Propagation of a femtosecond (fs) pulse train through an RQPM medium is illustrated in Fig. 2. Three-wave mixing via RQPM processes results in partial conversion of pulses to second and higher order optical harmonics. The noise spikes in the pulse train experience higher losses because the conversion process is nonlinearly dependent on intensity. As a result, the output pulse train has less intensity noise. Further, the RQMP process results in self-defocusing and self-phase modulation of pulses (SPM) due to the cascading of quadratic nonlinearities [19].

Thus, RQPM alters soliton dynamics in mode-locked oscillators via instantaneous nonlinear losses, self-defocusing, and SPM that balance self-focusing and SPM due to the Kerr nonlinearity of the gain medium. Another effect of RQPM, which is crucial in the context of spectroscopy, is nonlinear noise suppression. The influence of RQPM on the noise properties of mode-locked polycrystalline Cr:ZnS oscillators was studied in [20]. The noise reduction in ultrafast polycrystalline Cr:ZnS amplifiers will be illustrated in this section.

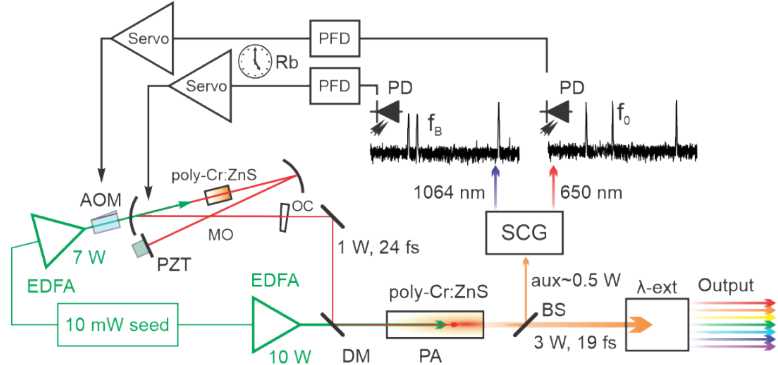

Fig. 3 Cr:ZnS optical frequency comb. MO, PA, polycrystalline Cr:ZnS master oscillator and power amplifier; EDFA, Er-doped fiber amplifiers; SCG, supercontinuum generation module; λ-ext, wavelength extension via nonlinear up- or down-conversion; PZT, piezo transducer; AOM, acousto-optic modulator; DM dichroic mirror; BS, beam splitter; PD, Si photodiode; PFD, phase-frequency detector; Rb, an atomic clock.

The Cr:ZnS comb source is built using the approach described in [21] and illustrated in Fig. 3. It consists of a polycrystalline Cr:ZnS mode-locked master oscillator (MO, fundamental wavelength 2.4 μm, average power 1 W, pulse width $\tau < 24$ fs, pulse repetition rate $f_R = 80$ MHz) and a single-pass polycrystalline Cr:ZnS power amplifier (PA, 3 W, $\tau \approx 19$ fs). MO and PA are optically pumped at the wavelength 1.567 μm by a purpose-developed ultra-low noise laser system, which includes two compact, Er-doped fiber amplifiers (EDFA) seeded by a common low-noise single-frequency semiconductor laser. All components of the system are water-cooled for acoustic noise reduction. A typical optical spectrum of the Cr:ZnS MOPA is shown in Fig. 4 (a).

For the comb referencing, about 20% of the combs' output is used to generate supercontinuum (SCG) in a bulk PPLN crystal [22]. The output SCG signal is band-pass filtered to 650 nm and 1064 nm sub-bands. The first sub-band is used for the detection of heterodyne beatings between the third and fourth harmonics of the fundamental comb at the combs' offset frequency $f_0$. The second sub-band is used for generation of a beat signal $f_B$ between a comb mode and a narrowband continuous wave laser acting as a reference, effectively providing a time base at the optical frequency. The rf signals $f_0$ and $f_B$ are then phase-locked to a Rb clock using a high-dynamic range servosystem. The typical timing jitter of the comb with respect to the optical reference is <0.1 fs. Further, the comb's control system includes a motorized stage for the adjustment of the pulse repetition frequency $f_R$ within 0.5 MHz and additional digital control loops for automatic compensation of the long-term drifts of the system and fluctuations of the environment. Thus, the comb can be robustly phase-locked to the time-base for an extended time.

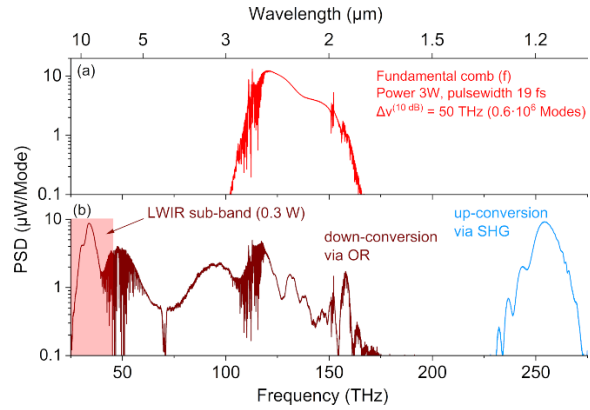

Fig. 4 (a) A typical spectrum of Cr:ZnS comb source at the fundamental wavelength. (b) Typical optical spectra obtained using optical rectification in ZGP crystal (OR) and second harmonic generation in PPLT crystal (SHG). The spectra are presented as power spectral density (PSD) of comb modes (log scale) vs mode's optical frequency. The PSD distributions are approximate and were derived from spectra measured at the axis of the laser beam with a Thorlabs OSA207C Fourier Transform optical spectrum analyzer

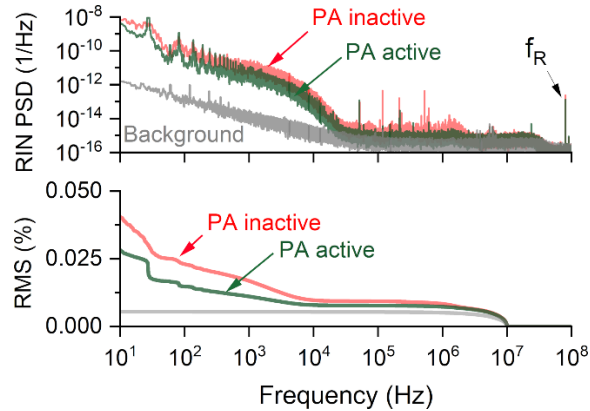

Fig. 5 (a) The RIN PSD of the Cr:ZnS MOPA. (b) Integrated RMS noise

For the extension of the fundamental comb to longer wavelengths, the Cr:ZnS MOPA is coupled to a ZGP crystal configured for optical rectification (OR) [23]. The typical power of an LWIR component in the output spectrum is about 0.3 W in the wavelength range 6.5 – 12 μm. Conversely, for up-conversion of the fundamental comb, we use second harmonic generation (SHG) in a PPLT crystal, generating a near-IR (NIR) comb with 1 W power. Typical optical spectra measured at the output of the LWIR and NIR extensions are illustrated in Fig. 4 (b).

The relative intensity noise (RIN) of the comb source was measured using a low-noise liquid nitrogen-cooled MCT photodetector having a 50 MHz bandwidth. The detected signal was digitized (16 bit resolution, 4 ns sampling interval) and post-processed on a PC. The top panel in Fig. 5 compares the RIN PSD of the Cr:ZnS MOPA at 1 W output power with the PA inactive and at full 3 W output power with the PA active. The noise peak in the RIN spectrum at the relaxation frequency ≈ 1 MHz is almost fully suppressed, in accordance with [20]. Remarkably, activation of the amplifier results in simultaneous amplification of pulses and in further noise suppression at all Fourier frequencies, which we attribute to the instantaneous nonlinear losses due to the RQPM in the amplifier's polycrystalline gain element. Furthermore, engaging the servo system does not produce any noticeable change in the RIN spectra. Thus, the integrated root-mean-square intensity noise of the phase-locked 3 W, 19 fs MIR comb source is, conservatively below 0.1% (see Fig. 5(b)). We consider these estimates preliminary, as our intensity noise measurements, especially at the Fourier frequencies above 1 MHz, are limited by the dynamic range of available photodetectors. We believe that the obtained noise level can be considered as ultralow – if not a record-low intensity noise – for a multi-Watt few-cycle optical frequency comb. We note that the most reliable evaluation of the laser source for spectroscopy is via the characterization of a spectrometer equipped with this source.

The phase noise of the Cr:ZnS comb was assessed using a recently proposed measurement technique that employs multi-heterodyne detection and subspace tracking [24]. The noise dynamics were found to be dominated by the phase noise component that is common to all comb modes, while all the other phase noise terms were below the measurement noise floor. Further, by analyzing heterodyne beatings of the comb with a stabilized ultra-narrowband laser (1.064 µm, linewidth <4 Hz, relative frequency stability <7·10$^{-15}$ in 1 s) we determined that the full width at half maximum (FWHM) linewidth of the comb modes is as low as 7.2 kHz for a free-running comb.

Thus, optical frequency combs based on ultrafast Cr:ZnS laser technology offer a set of features favorable for FTS and DCS. The high power and sub-3-cycle pulses of the comb greatly simplify its nonlinear down- and up- conversion to the desired spectral range. The low intensity noise of the comb enables the nominal, i.e., detector-limited, regime of spectroscopy. The ultra-low RIN PSD at high Fourier frequencies from 1 MHz to the $f_R$ = 80 MHz is especially beneficial for DCS. Last but not least, the very low phase noise and narrow linewidth of the comb modes greatly simplify robust mutual phase-locking of two combs, which is also an important requirement for a high-performance DCS.

## 3. EOS-DCS apparatus with parallel spectral acquisition

The spectrometer design is illustrated in Fig. 6. It is based on two Cr:ZnS combs, FC1 and FC2, at a 2.4 µm central wavelength and with mode spacings $f_{R1} \approx f_{R2} \approx$ 80 MHz, as described in the previous section. The combs are mutually phase-locked by locking their offset frequencies, $f_{01}$ and $f_{02}$, and phase-locking the combs' modes to the same optical reference at 282 THz and with fractional stability <7·10$^{-15}$.

For spectroscopy, the first comb (FC1) is down-converted to the LWIR (~0.3 W power at 6.5 – 12 µm) via OR and coupled to a 31.2 m multi-pass cell (Thorlabs HC30L-M02) with a gas sample. FC2 is up-converted to the NIR (1 W power at 1.2 µm) via SHG. The LWIR and NIR combs are then coupled to the EOS module [15, 25]. Here, the combs are sum-frequency-mixed, producing a new NIR signal, whose strength is proportional to the electric field of the LWIR comb. This field dependence is detected using ellipsometry, producing a periodic sequence of interferograms (IGMs) that coherently map the electric field of the LWIR comb to the rf domain with a large and adjustable mapping factor $f_{R1}/\Delta f_R$ where $\Delta f_R = f_{R1} - f_{R2}$ is the difference between the pulse repetition rates of two combs.

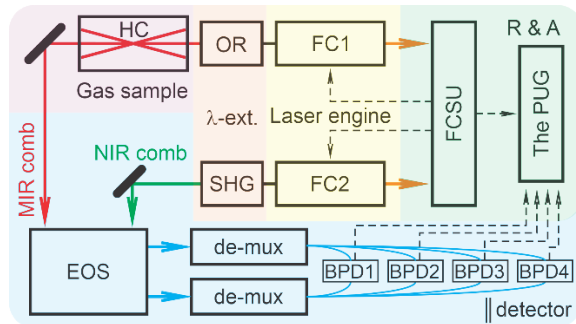

Fig. 6 EOS-DCS spectrometer consisting of a laser engine (FC1, FC2), frequency comb stabilization and data acquisition unit (FCSU + the PUG), wavelength extensions, gas sample in a Herriott cell, and a sub-system for generation, de-multiplexing, and parallel detection of multiple spectroscopy signals in the NIR, see text.

The EOS-DCS modality of Fourier transform spectroscopy thus allows accessing information about the electric field of a LWIR light source and, hence, complete information about the intensity and phase response of a gas sample with the low-cost, yet high-performance and versatile componentry provided by NIR photonics. This, in turn, enables LWIR Fourier transform spectroscopy with enhanced and scalable sensitivity. For instance, the specific detectivity of mass-produced low-cost miniature RT InGaAs photodiodes ($D^* \approx 2 \cdot 10^{12}$ Jones) is 1 – 2 orders of magnitude higher than that of much more expensive and bulky liquid nitrogen cooled or thermo-electrically cooled HgCdTe detectors, and 4 orders of magnitude higher than that of RT deuterated l-alanine triglycine sulfate (DLaTGS) detectors.

Higher detectivity of InGaAs photodiodes and, hence, higher dynamic range of the photodetection provides a proportional boost in the sensitivity of a LWIR Fourier spectrometer, as described in [8]. Further, the optical power of the NIR spectroscopy signals at the output of the EOS module (1 – 100 mW, depending on the configuration of the EOS-DCS apparatus) is sufficiently high to saturate a multitude of InGaAs photodiodes. This, in turn, provides additional enhancement in sensitivity via parallel acquisition. Importantly, de-multiplexing of the high-power NIR spectroscopy signal to the multitude of InGaAs photodiodes can be implemented using robust integrated photonics, e.g., fiber-optic WDMs or demultiplexers based on photonic integrated circuits.

To highlight the advantages of the EOS-DCS modality with parallel acquisition, we coupled the two NIR spectroscopy signals outputted by the EOS module to two single-mode fibers. We then used off-the-shelf fiber optic couplers to split copies of these NIR spectroscopy signals to a set of four balanced InGaAs detectors, producing four similar IGM signals. This symmetric regime of signal division potentially allows to achieve a square root dependence of the sensitivity on the number of photodetectors $N_D$. However, a linear dependence of the sensitivity on $N_D$ can also be achieved by de-multiplexing with spectral filtering, so each detector measures a separate sub-band of the whole NIR spectroscopy signal [8].

Next, the four spectroscopy signals $IGM_{1,2,3,4}$ and four signals $f_{01}$, $f_{02}$, $f_{B1}$, $f_{B2}$ from the comb stabilization unit (FCSU) are digitized (14 bit resolution, 8 ns sampling interval) and fed to a sub-system for real-time data processing (the PUG, [26]). The digital signal processing implemented by the PUG includes:
(i) Retrieval of the information about the residual phase noise and timing jitter of the combs from the reference signals $f_{01}$, $f_{02}$, $f_{B1}$, $f_{B2}$.
(ii) Fast digital corrections of IGM data streams using the retrieved phase and timing error signals
(iii) Additional correction of IGM data streams to compensate for out-of-loop drifts of the interferometer
(iv) Re-sampling, alignment, and averaging of IGMs, Retrieval of the optical axis from the reference frequencies $f_{01}$, $f_{02}$, $f_{B1}$, $f_{B2}$.

Thus, the initial spectroscopy data stream (1.5 Gb/s per acquisition channel) is converted to a corrected and averaged IGM data stream (30 MB per acquisition channel per averaging period) that can be stored to the computer memory at a desired rate for further analysis.

## 4. EOS-DCS figure of merit, stability, and detection sensitivity

A typical time domain IGM signal produced by our EOS-DCS apparatus is illustrated in Fig. 7. Information about the electric field of the LWIR pulses and, hence, about the source's spectral envelope and phase, is encoded in the IGM's main peak. The intensity and phase response of a gas sample is encoded in the IGM's weak, extended tail, the so-called free induction decay (fid, Fig. 7 (c)). The IGM signal also contains responses from optical interfaces along the beam path. This information is valuable for, e.g., optical coherence tomography. However, in the case of spectroscopy, the reflections on interfaces, so-called etalons, are undesirable because etalon signals couple to the main fid signal. In our set-up, the beam path was designed to minimize such reflections on interfaces; the residual etalons visible in Fig. 7 (c) are overspill etalons due to the spatial overlap between beam passes in the Herriott cell and were digitally removed from the signal during post-processing.

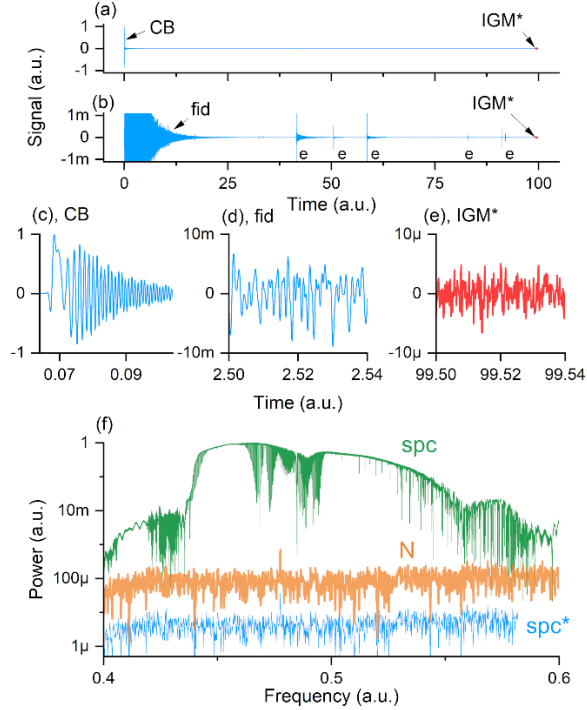

Fig. 7 Corrected and averaged IGM consisting of 1.9 million temporal samples shown in time and frequency domains. (a) and (b) full IGM signal and the same signal shown with 1000× vertical zoom: CB, center burst at 0 time delay; fid, free-induction decay; e, etalons; IGM* a part of the IGM at 99% time delay that is used for the noise measurements. (c), (d), and (e) parts of the IGM signal showing details of the CB, fid, and IGM*, respectively (note 100× vertical zooms in each consecutive plot). (f) spectral view of the IGM signal: spc, spc* are power spectra of the full IGM and IGM*, respectively. N is the noise power spectrum obtained from spc* by its normalization to the full length of the time window.

The convenient metric that facilitates the evaluations of performance and comparisons between different spectrometers is the figure of merit. The authors of [8] introduced DCS FOM as a product of the number of resolved comb modes M (i.e., the number of spectral elements measured in parallel) and the spectral SNR attained per unit time √t.

$$FOM_{BOX} = M \frac{SNR_{spc}}{\sqrt{t}} \qquad (1)$$

This definition is very useful for theoretical studies and qualitative comparisons; however, it is unambiguous only for boxcar-shaped optical spectra, which is never the case in real life. Therefore, the values reported in the literature are difficult to compare. Throughout this paper, we will use an equivalent but spectral-shape and baseline independent definition introduced in [27]:

$$FOM_\Sigma = \frac{1}{\sqrt{t}} \frac{\sum |S_k|}{\sigma_{spc}} = \frac{1}{\sqrt{t}} \Sigma SNR \qquad (2)$$

where the sum of the spectral elements in the PSD is the total spectral power, and $\sigma_{spc}$ is the spectral noise standard deviation. Further, to make our computations of the FOM baseline independent, we calculated $\sigma_{spc}$ from the fast Fourier transform of a small part of the IGM signal that does not contain any fid or etalon signal, shown in Fig. 7 as IGM* at the tail of the IGM signal.

The FOM of our EOS-DCS apparatus was measured and its stability was assessed by measuring and analyzing the dependencies of $\Sigma_{SNR}$ on the square root of the acquisition time $\sqrt{t}$. The linear part of a $\Sigma_{SNR}$ vs $\sqrt{t}$ dependence corresponds to the effective time of coherent averaging $\tau_{coh}$, and the slope of this linear trend is the $FOM_\Sigma$ of the instrument. Experimental results are summarized in Fig. 8 and compared with the data collected from a high-end FTS spectrometer [13]. Since in both cases we used essentially the same high-power, low-noise Cr:ZnS-based LWIR sources, we can evaluate the limitations imposed by the instruments rather than the sources. Although two experiments were carried out at different spectral resolutions (80 MHz for EOS-DCS vs 3 GHz for FTS) and in different spectral windows (800–1400 cm$^{-1}$ for EOS-DCS vs 850–3500 cm$^{-1}$ for FTS), those differences are captured nicely by our definition of the FOM.

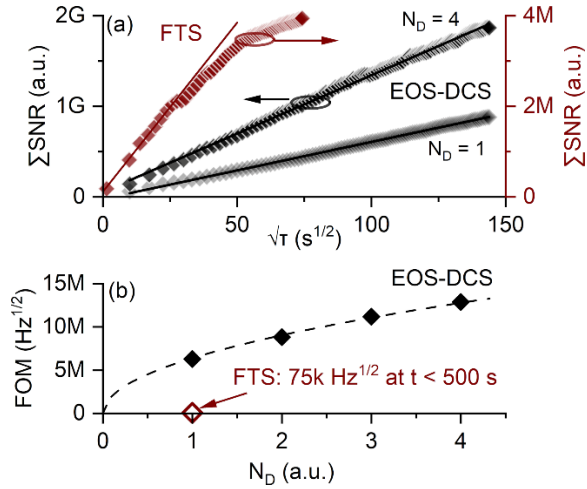

Fig. 8 (a) Measured $\Sigma_{SNR}$ vs $\sqrt{t}$ dependencies for EOS-DCS with a single detector ($N_D = 1$) and with four detectors ($N_D = 4$) and for FTS ([13], right vertical axis) are compared. (b) solid symbols ♦ Measured dependence of the EOS-DCS FOM vs ND; dashed line square root fit of the measurement; open symbol ◊, the FTS FOM achievable if the acquisition time is less than 500 seconds.

In the case of a single detector ($N_D = 1$), the EOS-DCS instrument features 84 higher FOM in comparison to the FTS instrument. This significant enhancement in the spectroscopic performance highlights two main advantages of the EOS-DCS approach: (i) much higher resolution and, hence, larger number of spectral elements measured in parallel (ii) higher dynamic range of the LWIR photodetection via the EOS method. The achieved advantage in the FOM also confirms ultra-low intensity and phase noise of the Cr:ZnS combs: the noise in the system is dominated by the detector noise rather than the light source noise, which, in turn, allows to measure more spectral elements in parallel with a single detector without compromising the spectral SNR.

Figure 8(a) also highlights the advantage of EOS-DCS in terms of stability. In the FTS case, the $\Sigma_{SNR}$ vs $\sqrt{t}$ plot starts to deviate from the linear trend after about $5 \cdot 10^2$ seconds. On the other hand, the trends for DCS remain perfectly linear and, hence, the averaging is coherent for the whole duration of the measurement ($2 \cdot 10^4$ seconds in the particular case of Fig. 8).

Further, parallel spectral acquisition with multiple detectors provides an additional boost to the spectroscopic performance. Figure 8 (b) illustrates the FOM dependence on the number of detectors $N_D$. As can be seen, the experimental data matches the square root dependence that is a best-case scenario for the regime of symmetric signal division. As a result, the EOS-DCS

FOM at $N_D = 4$ is 160 times higher than that for FTS. Importantly, since we implemented the parallel spectral acquisition in the NIR, the above boost in performance was achieved at almost no additional cost or increased complexity of the system.

To compare the detection sensitivity of the EOS-DCS and FTS instruments, we have assessed the Allan-Werle deviations (AWD) for retrieved concentrations of ethylene ($C_2H_4$) and ammonia ($NH_3$). In both cases, we evaluated these gases for an $N_2$ buffer and using the same 31.2 m long multi-pass cell with a pressure of 100 mbar (ethylene) and 1000 mbar (ammonia). The EOS-DCS measurements were carried out at a native 80 MHz resolution and in the spectral window 800–1400 cm$^{-1}$. During the FTS measurements, the resolution was set to 3 GHz and the spectral window to 900–1250 cm$^{-1}$. The results are summarized in Fig. 9. For ethylene, we obtained a 96 ppb·Hz$^{1/2}$ trend for EOS-DCS with a single detector compared with a 1.68 ppm·Hz$^{1/2}$ trend for FTS. Thus, we achieved a factor of 18 improvement in the sensitivity at 37.7 times higher resolution and in a 1.7 times broader spectral window. Using parallel acquisition, the sensitivity was enhanced by another factor of two, yielding the 48 ppb·Hz$^{1/2}$ dependency that matches the theoretical limit for the symmetric division to four photodetectors. Further, the AWD plot for EOS-DCS follows the linear trend over an extended time, achieving a 0.48 ppb detection limit in $10^4$ seconds. Thus, in terms of stability, we achieved a 12-fold improvement in the optimal averaging time and, consequently, a 170-fold improvement in the detection limit.

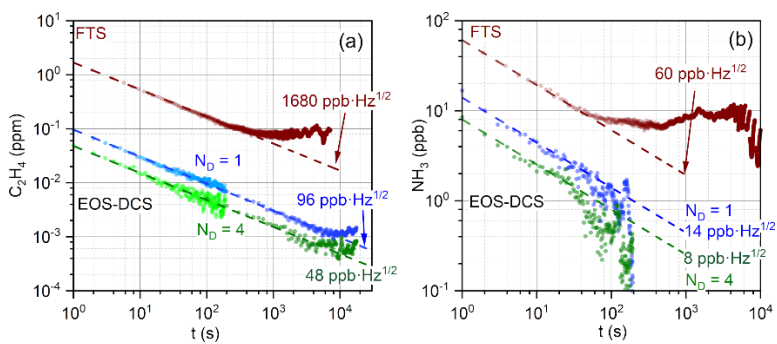

Fig. 9 Allan-Werle deviations for ethylene (a) and ammonia (b). The results obtained using FTS and EOS-DCS are compared. Dashed lines show the t$^{-1/2}$ dependencies

For ammonia, we obtained 14 ppb·Hz$^{1/2}$ and 8 ppb·Hz$^{1/2}$ dependencies using EOS-DCS with a single detector and four detectors, respectively. We conservatively estimate the detection limit as 0.76 ppb of ammonia in 100 s – an order of magnitude improvement in comparison to FTS. In the case of EOS-DCS, the estimation of the detection limit at longer timescales was hindered by instabilities in the gas sample: fluctuations in the ambient temperature and pressure in the lab resulted in fluctuations of the $NH_3$ concentration by 1 – 5 ppb per minute. While it is encouraging that our EOS-DCS instrument has the sensitivity to resolve those fluctuations, the instabilities of the sample have resulted in divergence of AWD plots at t > $10^2$ seconds.

Thus, our experiments demonstrate that our EOS-DCS apparatus with parallel acquisition outperforms state-of-the-art FTS instruments on all accounts, simultaneously providing 1 – 2 orders of magnitude higher sensitivity at much higher resolution and in a broader spectral window. The above combination of features is uniquely suited for fast quantitative analysis of complex gas mixtures containing different molecules at vastly different concentrations.

## 5. Quantitative analysis of a complex gas mixture

To demonstrate the capabilities of our EOS-DCS instrument for practical applications, we analyzed a custom gas mixture that imitates human breath. The original mixture consisted of 2 ppm of methane ($CH_4$), 1 ppm of each methanol ($CH_3OH$), isoprene ($C_5H_8$), acetone (($CH_3)_2CO$), and 51000 ppm carbon dioxide ($CO_2$), diluted in $N_2$ buffer gas. During the loading of the mixture into the multi-pass cell, a small amount of water vapor (340 ppm) entered the cell as well. As the mixture remained in the cell for several hours, we observed significant changes in gas concentrations, which we attribute to aggregation of molecules on the walls of the cell and in the connecting tubing. Acetone in particular is known to adsorb readily onto surfaces, and its concentration is expected to drop significantly during this period. We also noted the appearance of trace amounts of $NH_3$, which, most likely, has aggregated in the cell during previous measurements. Figure 10 illustrates the IGM and raw spectrum obtained after $10^3$ seconds of averaging, together with the fully post-processed spectrum, which included the optical axis retrieval, etalon removal, baseline correction, and conversion to absorbance units. The spectrum shown in Fig. 10 (c) is dominated by strong absorption lines of water vapor and $CO_2$.

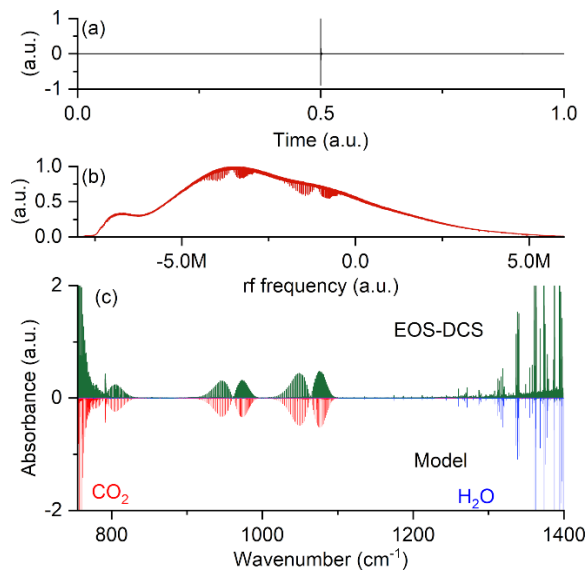

Fig. 10 LWIR spectrum of a custom gas mixture at 72 hPa acquired in $10^3$ seconds: (a) averaged IGM; (b) FFT of the IGM; (c) retrieved absorbance spectrum of the sample.

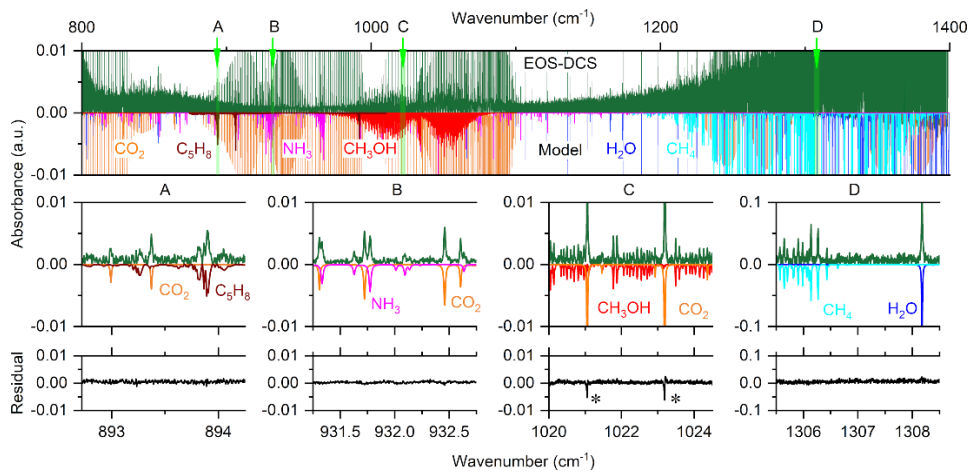

Fig. 11 Top: absorbance spectrum of the sample shown with 200× zoom along the absorbance axis); light-green rectangles marked as A, B, C, D indicate details of the spectrum. Middle: details of the spectrum shown in high spectral resolution (300× zoom along frequency axis), revealing absorption lines of isoprene at 0.7 ppm (A), ammonia at 0.065 ppm (B), methanol at 0.7 ppm (C), and methane at 2 ppm (D). Bottom: The residuals of the fits. *, Increased noise in the residual at the absorption peaks of $CO_2$ is caused by saturation of these strong absorbances.

The next figure shows details of the same spectrum at low levels of absorbance and at high resolution. As can be seen, the quality and dynamic range of the spectrum obtained in $10^3$ seconds are high enough to identify and quantify the concentrations of all the molecules in the mixture, except acetone. Specifically, we measured 2 ppm for methane, 65 ppb for ammonia, 0.7 ppm for methanol, and 0.7 ppm for isoprene. Most likely, the concentration of acetone in the cell is below 0.5 ppm due to particularly strong aggregation of this molecule on the walls of the cell.

## 6. Conclusion

Since its introduction about 20 years ago, dual comb spectroscopy has been praised for its high resolution, precision, and a high speed at which the individual spectra can be acquired. Here we show that it can also deliver sensitivity far beyond the reach of any direct-detection LWIR Fourier spectrometer, without compromising high resolution and broad bandwidth.

This performance is enabled by bridging the LWIR spectroscopy and NIR photonics. By combining high-power, low-noise ultrafast driving lasers with DCS technique and electro-optic sampling detection (EOS-DCS), we map the full amplitude and phase response of a LWIR spectrum onto a NIR signal. The 80 MHz spectral resolution – much higher than that for most conventional FTS instruments – enables resolution of closely spaced absorption features essential for unambiguous species identification. Using a single balanced InGaAs photo detector, we demonstrated sub-ppb sensitivity with simultaneous coverage of multiple molecular species in a single measurement, beyond the reach of any direct-detection LWIR Fourier spectrometer reported to date.

The near-infrared detection architecture further enables parallel spectral acquisition with minimal additional system complexity. Employing four balanced InGaAs detectors provides an additional twofold improvement in sensitivity, consistent with symmetric signal division. Owing to the high optical power of the electro-optically sampled signal, this approach supports straightforward scaling of sensitivity through increased detector count. With proper de-multiplexing it enables sensitivity scaling linearly with the number of detectors, establishing a clear pathway toward broadband LWIR spectroscopy with sub-ppb sensitivity at high acquisition rates and ppt-level detection on minute timescales.

As a proof-of-concept, we successfully identified and quantified multiple trace molecules in a synthetic breath-like mixture, demonstrating the potential of EOS-DCS for complex gas analysis. This new performance envelope establishes EOS-DCS as a compelling platform for next-generation broadband trace gas spectroscopy, combining sub-ppb sensitivity, high-resolution molecular fingerprinting, and simultaneous multispecies detection in a single measurement.

## 7. Back matter

**Funding**

R.K., and S.C. acknowledge funding by European Partnership on Innovative SMEs (Eurostars) GreenFruit (ID: 4766); D.K, A.M., and K.V. acknowledge funding by US Office of Naval Research (ONR) award N00014-18-1-2176, US Air Force Office of Scientific Research (AFOSR) awards FA9550-23-1-0126 and FA9550-24-1-0196, U.S. Department of Energy, award DE-SC0012704.

**Acknowledgments**

S.V., I.M., and M.M. thank the scientific and engineering teams at IPG Photonics for support of this project.

**Disclosures** The authors declare no conflicts of interest.

**Data availability**

Data underlying the results presented in this paper are not publicly available at this time but may be obtained from the authors upon reasonable request.